\documentclass[review,12pt]{elsarticle}

\usepackage{amssymb,amsmath}
\usepackage{array}
\usepackage{multirow}
\usepackage{multicol}
\usepackage{tabularx}
\usepackage{colortbl}
\usepackage{nicefrac}
\usepackage{mathrsfs}
\usepackage{verbatim}
\usepackage[T1]{fontenc}

\usepackage[squaren, binary]{SIunits}

\graphicspath{{.},{./}}

\usepackage{lineno}

\biboptions{compress}

\journal{Nuclear Instruments and Methods in Physics Research Section A}

\begin{document}

\begin{frontmatter}



\title{Simulation study on light ions identification methods for carbon beams from 95 to 400~MeV/A}


\author{S.~Salvador\corref{cor1}}
\ead{salvador@lpccaen.in2p3.fr}

\author{M.~Labalme, J.M.~Fontbonne, J. Dudouet, J.~Colin, D.~Cussol\corref{}}

\address{Laboratoire de physique corpusculaire de Caen, ENSICAEN, 6 boulevard du Mar\'echal Juin, 14050 Caen cedex, France}
\cortext[cor1]{Corresponding author. Tel.: +33 2 31 45 25 54}

\begin{abstract}

Monte Carlo simulations have been performed in order to evaluate the efficiencies of several light ions identification techniques. The detection system was composed with layers of scintillating material to measure either the deposited energy or the time-of-flight of ions produced by nuclear reactions between $^{12}$C projectiles and a PMMA target. Well known techniques such as $\Delta E$---Range, $\Delta E$---$E$---ToF and $\Delta E$---$E$ are presented and their particle identification efficiencies are compared one to another regarding the generated charge and mass of the particle to be identified. The simulations allowed to change the beam energy matching the ones proposed in an hadron therapy facility, namely from 95 to 400~MeV/A.

\end{abstract}

\begin{keyword}
Hadron therapy \sep Particle Identification \sep Monte Carlo Simulations


\end{keyword}

\end{frontmatter}

 \linenumbers

\section{Introduction}

Particle identification is of major importance in multiple fundamental physics experiments and especially for nuclear reaction studies. Various methods can be used, mostly based on the Bethe-Bloch formula, to retrieve either the partial energy $\Delta E$, lost in a thin detector, the total energy $E$ in a thick one~\cite{Dudouet201398}, the $\beta$ parameter using the particle velocity based on time-of-flight (ToF) measurements~\cite{Bass1975125}, or the range~\cite{Chulick1973171, Greiner1972291, Amaldi2011337} as well as the Bragg peak amplitude~\cite{Asselineau1982109}. The associated detection system can be made of solid state detectors, such as germanium or silicon allowing very good estimation of the deposited energy and pulse shape discrimination~\cite{Mutterer,LeNeindre}; scintillating material, either organic or inorganic, for good timing resolution particularly in high energy physics~\cite{Banerjee1988121}; or gaseous detectors as a low density stopping medium~\cite{Strittmatter} for low kinetic 
energy ions. The detection system is then designed and optimized for the purpose of the technique used. It is thereby difficult to know {\it a-priori} the most efficient method for identification when designing a new experiment.

%
%

In this paper, we have performed simulation studies for particle identification in multi-fragmentation processes of carbon beams with targets at energies ranging from 95 to 400~MeV/A. We focused on three methods, based on $\Delta E$---Range, $\Delta E$---$E$---ToF and $\Delta E$---$E$ measurements done by scintillating detectors only. Solid state and gaseous detectors have been left aside due to their poor timing resolution ($>$1~ns) and too low density ($\sim$1~mg~cm$^{-3}$), respectively.
The detection system is the same for all measurements (save for the thickness of the thin stage) to be able to compare identification techniques and not the system performances.

The goal of this work is to investigate an efficient method able to discriminate 1 atomic mass up to $^{12}$C ions. The system will be used in double differential cross-section measurement experiments for carbon therapy interest, at the new Advanced Resource Center for HADron therapy in Europe (ARCHADE) based in Caen.

\section{Simulation materials and methods}
\label{MatandMeth}
The simulations were based on the GEANT4 Monte-Carlo toolkit~\cite{Agostinelli2003250}. The GEANT4 version used is the 9.5 with the physics list QMD (Quantum Molecular Dynamics) for inelastic reactions associated with an FBU (Fermi Break-Up) de-excitation process. This physics list has been chosen instead of the current BIC (Binary Intra-nuclear Cascade) package due to its cross-sections of fragments production closer to experiments, particularly for energy distributions~\cite{deNapoli}.

The simulations consisted on the interaction of 10$^6$ $^{12}$C ions at normal incidences with a spherical PMMA target of 5~mm in diameter performed in ultra vacuum. For each event, the interactions of the primary particle or secondaries with the detection system were recorded. For each ion, its Z and mass value (A) are known and compared in the post processing analysis to the measured one using different identification method, described in the following sections. The system can detect events coming from the beam which have not encountered any fragmentation processes in the target. These events are the most likely ones. However, to avoid degradation of the results, data from primary ions, i.e. encountering no inelastic processes in the target, were not been used in the identification processes. This will be discussed separately in the appropriate section.

Multiple simulations were done by changing the beam energy from 95~MeV/A (maximum energy provided by GANIL in Caen), to 200, 300 and 400~MeV/A, representing appropriate energies for carbon therapy purposes.
~\\

The simulated detection system was based on thallium doped cesium iodide scintillating crystals (CsI:Tl) with a density of 4.51 g cm$^{-3}$, a decay time of 1~$\mu$s and a light yield of 
$\sim$55~ph~keV$^{-1}$~\cite{dehaas}. This crystal has been chosen due to its known quenching factors, that allowed the conversion of the deposited energy into scintillation light for better accuracy. This conversion was made according to the formula given in~\cite{Parlog2002693}: 

\begin{eqnarray}
 L & = & a_1\Bigg\{E_0\left[1-a_2\frac{AZ^2}{E_0} \ln\left(1+\frac{E_0}{a_2AZ^2}\right)\right] \nonumber\\
& &+a_2a_4AZ^2 \ln\left(\frac{E_0+a_2AZ^2}{a_3A +a_2AZ^2}\right)\Bigg\},
\end{eqnarray}

where $L$ is the scintillation light in equivalent number of photoelectrons, $E_0$, the deposited energy in keV, $a_1$, the conversion factor from energy to converted photoelectrons, $a_{\{2...4\}}$ are the quenching factors, $A$ and $Z$, the mass and atomic number of the ion. Table~\ref{quenchingFactors} gives a summary of the quenching factors while $a_1$ represents in our case the light yield of the scintillator times the photon detection efficiency of the associated photodetectors. The photon detection efficiency is taken as the quantum efficiency ($\varepsilon_q=0.25$) of the photodetector such as a photomultiplier tube, times the collection efficiency taken to be around 50\%. 
In the following sections, energy will always be expressed as the measured output light, in terms of photoelectrons, even if mentioned as energy.

\begin{table}[!ht]
\caption{Values of the quenching factors used in the simulations~\cite{Parlog2002693}.}
\centering
\footnotesize
\label{quenchingFactors}
\renewcommand{\arraystretch}{1.3}
{
\begin{tabularx}{\linewidth}{XXXX}
\hline\hline
$a_1$ &  $a_2$ &$a_3$ &$a_4$ \\ \hline
6875 & 0.71 & 3.8 & 0.26    \\
\hline\hline
\end{tabularx}}
\end{table}

To introduce the detector energy resolution, each amount of converted photoelectrons values were randomly extracted from a gaussian distribution with $L$ as mean value and sigma given by :
\begin{eqnarray}
 \sigma_L = \frac{L}{2.35}\times \left( \frac{1.021}{\sqrt{E_0}}+0.019\right).
\end{eqnarray}

The energy resolution parameters were derived from experimental energy resolutions given by~\cite{Mares2004353}.

The detector was composed of scintillating layers of 120~$\times$~120~cm$^2$ and increasing thicknesses. Each layer scaled with depth from 0.2~mm to 13~mm thick by 0.2~mm steps to accurately sample the small ranges and to be able to optimize the thickness of the $\Delta E$ stage. Using 65 layers, the total depth of the detector is 42.9~cm allowing to entirely stop protons up to 480 MeV. 
The thickness of the first stage is optimized by finding the minimum value of number of layers which minimizes the identification errors. This basically means that one need to maximize the deposited energy while minimizing the number of inelastic interactions inside the corresponding material thickness. Table~\ref{thicknesses} gives a summary of the optimized thicknesses (the sum of layers thicknesses considered for the $\Delta E$ stage) used at the different energies for the $\Delta E$---ToF and $\Delta E$---$E$ methods.

\begin{figure}[!ht]
\includegraphics[clip=true, trim=0. 0. 0. 0.cm, width=\linewidth]{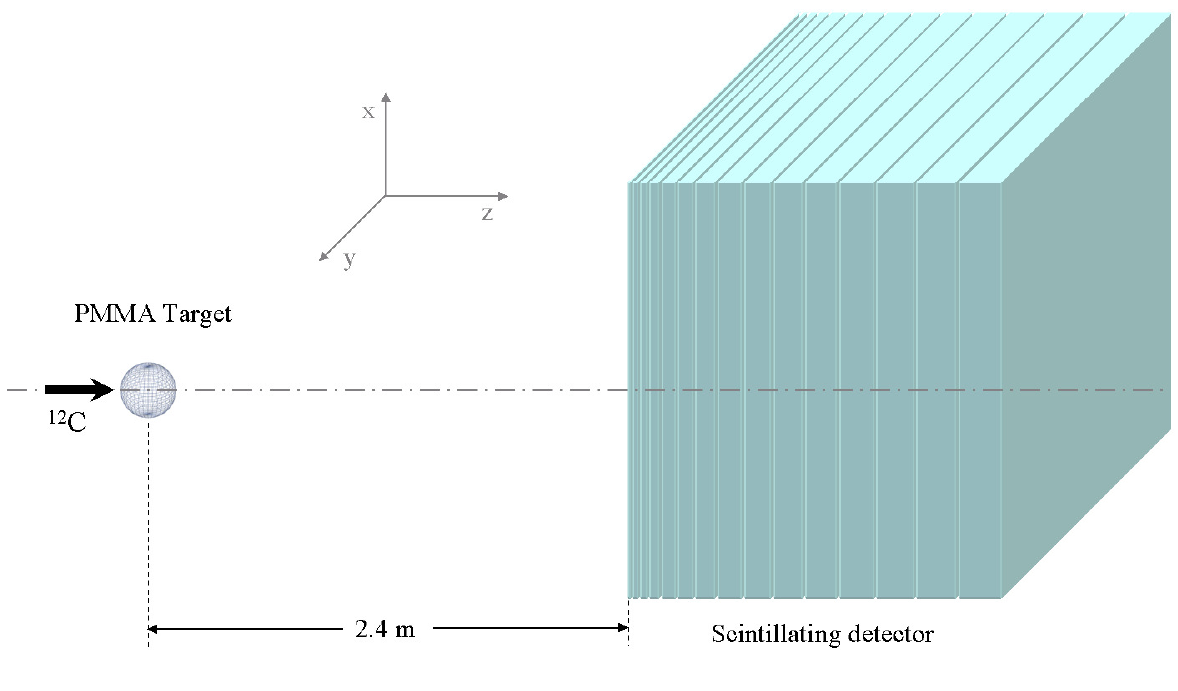}
\centering
\caption{Schematic view of the detection system. Not at scale for clarity reasons.}
\label{dispositif}
\end{figure}

The system was located at 2.4 m from the target offering a $\pm$13$\degree$ opening angle and a good ToF measurement. Fig.~\ref{dispositif} gives a schematic view of the simulation set-up.

\begin{table}[!ht]
\caption{Thickness of the $\Delta E$ layer in mm at different beam energies.}
\centering
\footnotesize
\label{thicknesses}
\renewcommand{\arraystretch}{1.2}
{
\begin{tabularx}{\linewidth}{XXX}
\hline\hline
Beam energy &  \multicolumn{2}{c}{Thickness (mm)}  \\\cline{2-3}
(MeV/A)     &  $\Delta E$---ToF method & $\Delta E$---$E$ method  \\ \hline
95 & 0.2 &   0.6  \\
200 & 0.6 &   1.2  \\
300 & 0.6 &   2.0 \\
400 & 2.0 &   2.0  \\
\hline\hline
\end{tabularx}}
\end{table}

Despite the fact that ions can easily be tracked in depth (in the $z$ direction), tracking in the $xy$ plane was not used except to distinguish the energy deposition by each individual ion. It is obvious that such a system is very unlikely to be built. First of all, the thicknesses of the layers, particularly at small ranges, should be increased to allow the fabrication process and the use of appropriate photodetectors. Then, each layer would be individually divided into small tills to have some tracking information in the $xy$ plane. Some detection systems based on the same principle have already been tested for hadronic granular calorimetry dedicated to particle physics~\cite{Andreev2005368}.

The following paragraphs will describe the techniques used for particle identification.

\subsection{The $E$---Range method}

The $E$---Range method is usually used for identification of particles with ranges measured in a gaseous detector within few tens of centimeters. Here, the method is 
presented to identify charged particles with much higher velocities detected in a dense material (density of 4.51 g cm$^{-3}$). 

The relation between the energy and the range (eq.~\ref{eqE-Range}) has been derived from the Bethe-Bloch formula by Greiner~\cite{Greiner1972291} for $\beta$ values under 0.7 which correspond approx. to 370 MeV/A.
\begin{eqnarray}
\label{eqE-Range}
 E=a_1A\left(\frac{bRZ^2}{A}\right)^c,
\end{eqnarray}
where $a_1$ is the conversion factor for energy to photoelectrons, $b$ and $c$ are fit parameters and $R$ is the range.

By measuring simultaneously the total deposited energy and the range of an unknown ion, its charge and mass can be obtained.

\subsection{The $\Delta E$---$E$---ToF method}

While the $\Delta E$---Time-of-Flight method is used to measure the charge of ions, the Energy---ToF method can be used to obtain their mass.
Using the Bethe-Bloch formula to obtain the charge dependence of the deposited energy in a $\Delta x$ thin medium (eq.~\ref{eqdETof}) and the relativistic equation of a particle total energy (eq.~\ref{eqETof}), one can adjust fit parameters to identify the ions in two different $\Delta E$---ToF and $E$---ToF distributions.

\begin{eqnarray}
\label{eqdETof}
 \Delta E & = & a_1\frac{Z^2}{\beta^2}\left[ \ln\left(\frac{\beta^2 b_{mat}}{1-\beta^2} \right)- \beta^2 \right]\Delta x,
 \end{eqnarray}
\begin{eqnarray}
\label{eqETof}
E&=& a_1 b_{\{1...12\}} u A \left( \frac{1}{\sqrt{1-\beta^2}}-1 \right),
\end{eqnarray}

where $b_{mat}$ is a fit parameter depending on the detector material, $\beta = \frac{d}{c\times ToF}$ with $d$, the distance to the detector, $c$, the speed of light, and $u$ is the unified atomic mass of 931.494 MeV~c$^{-2}$. This approximation stands due to the use of correction parameters $b_{\{1...12\}}$ obtained from the fit for each ion mass.

To introduce a measurement coincidence time resolution, ToF measurements are obtained from random values of a Gauss distribution where the full width at half maximum (FWHM) has been set to 300~ps. This value, even if not achievable using CsI:Tl crystals, can be measured with good timing detectors using fast scintillators and photomultiplier tubes~\cite{Moszynski200631}.
~\\

One main disadvantage of these two techniques is that they relate independently to the same particle due to the need of two different plots for identification. This leads to some very unlikely isotopes identification coming from a mass value uncorrelated to a Z value. As a consequence, the given results only take into account isotopes supposed to be produced by the initial reaction between $^{12}$C and nuclei in the target.

\subsection{The $\Delta E$---$E$ method}

The $\Delta E$---$E$ method is often used to identify charged particles even with energies up to few hundred MeV/A whether using gaseous, solid state or scintillating detectors for both measurements of $\Delta E$ or $E$. It relies on the detection of the energy deposited by particles in a thin detector as a function of the residual deposited energy in a sufficiently thick detector to stop the 
particle.

A usual functional of the relation between $\Delta E$ and the residual energy is given in~\cite{TassanGot} by:

\begin{eqnarray}
 \Delta E & = &\Big[\left(gE\right)^{1+\mu} + (\lambda Z^{\alpha}A^{\beta})^{1+\mu} \nonumber \\
 &&+ \xi Z^2A^{\mu}(gE)\Big]^{\frac{1}{1+\mu}}-gE,
\end{eqnarray}

with $g$, $\mu$, $\lambda$, $\alpha$, $\beta$ and $\xi$ are parameters obtained by fitting the distributions for each couple (Z, A). The parameter $\lambda$ includes the $a_1$ parameter as well as the thickness of the first stage, $\Delta x$.

\subsection{Particle identification}

Using the different analytical solutions given by the equations, identification of a particle was made for each method by a Newton-Raphson approach. In this case, the distance of an event to 
the curve was minimized in few steps, making the event to be on the normal to the curve's tangent.
An event was then attributed to a curve for the smallest event-to-curve distance when testing for all curves, relating the event to a particular Z and/or A value.

To compare the identification efficiency with the known ion charge and mass, distributions of the charge and the particle identification parameter (PID, taken as 0.8$\times$Z+0.1$\times$A\footnote{For instance, tritons have a PID equal to 1.1 and $\alpha$ particles have a $\textrm{PID}=2.0$.}) were built for both measured and real values. The number of measured counts at a particular PID ($N_{meas}$) was then compared to the corresponding one in the generated distribution, $N_{true}$. The result was normalized by $N_{true}$ to obtain the relative identification error (RIE, eq.~\ref{eqIdError}). As a result, this method took into account all sources of identification errors but is also dependent on the isotope statistic.

\begin{eqnarray}
\label{eqIdError}
 RIE=\frac{\textbar N_{meas} - N_{true}\textbar}{N_{true}}.
\end{eqnarray}

\subsection{Energy evaluation}

For each method, the measured energy of a {\bf well identified} particle (PID$_{meas}$=PID$_{true}$) is compared to the generated one. The energy can then be obtained either by the total deposited energy (or the sum of the partial and residual energy) or by time-of-flight measurements. Special care was taken when evaluating the energy by the ToF method. This one was obtained using a similar formula as eq.~\ref{eqETof}, except that the Z and A of the particles are known and that the proton and neutron masses as well as the binding energies can be used instead of the parameters $b_{1...12}$. The energy in MeV is then converted in photoelectrons using the parameter $a_1$.

A plot is thereafter made summing all detected particles. The ratio of ions with a measured energy truncated due to losses after inelastic collisions, $R_ {trunc}$, is extracted. First, the FWHM of the peak centered around 1 is evaluated. Assuming a gaussian distribution, a lower limit of -5$\sigma$ (using $1\sigma=\textrm{FWHM}/2.35$) to the position of the peak is obtained. For each method and beam energy, the ratio is measured by summing the number of events up to the limit divided by the total amount of events in the distribution. This parameter evaluates then the effect of the method in the  degradation of the ion energy even if this one has been well identified.


\section{Results}
\subsection{Particle identification}

\begin{figure}[!ht]
\includegraphics[clip=true, trim=0.cm 0. 0.2cm 0.75cm, width=\linewidth]{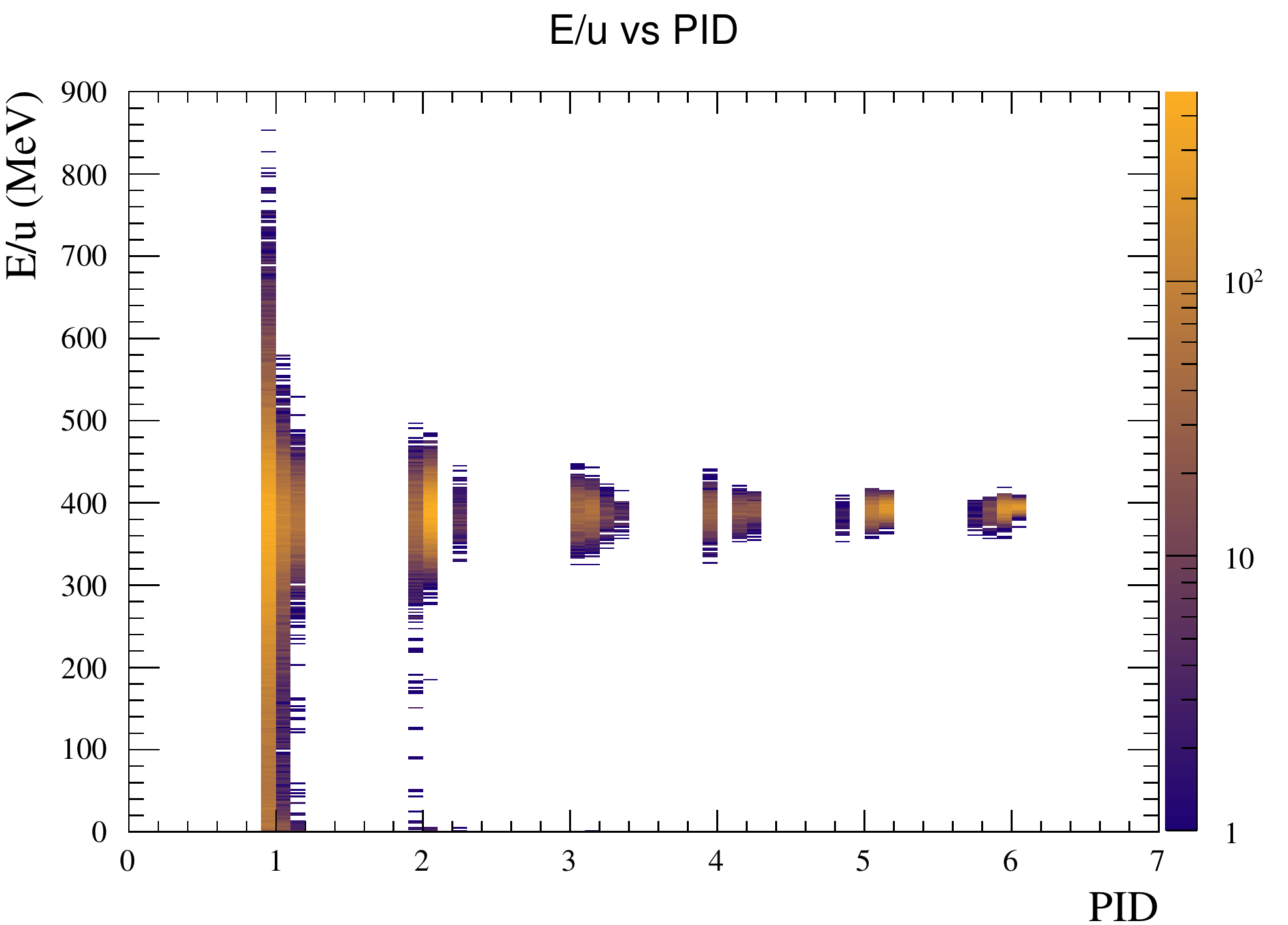}
\centering
\caption{Energy distribution of the different detected isotopes for $\textrm{E}_{beam}=400~\textrm{MeV/A}$.}
\label{figEnergy400MeV}
\end{figure}

Fig.~\ref{figEnergy400MeV} shows an example of the energy per nucleon distributions of the isotopes generated by the simulation and detected by the system for $\textrm{E}_{beam}=400~\textrm{MeV/A}$.
At this energy, the distributions are well centered around the beam energy representing mostly fragments from the projectile, except for protons ($\textrm{PID}=0.9$) and deuterons ($\textrm{PID}=1.0$) which exhibit broader distributions.

\begin{figure*}[!ht]
\includegraphics[width=\linewidth]{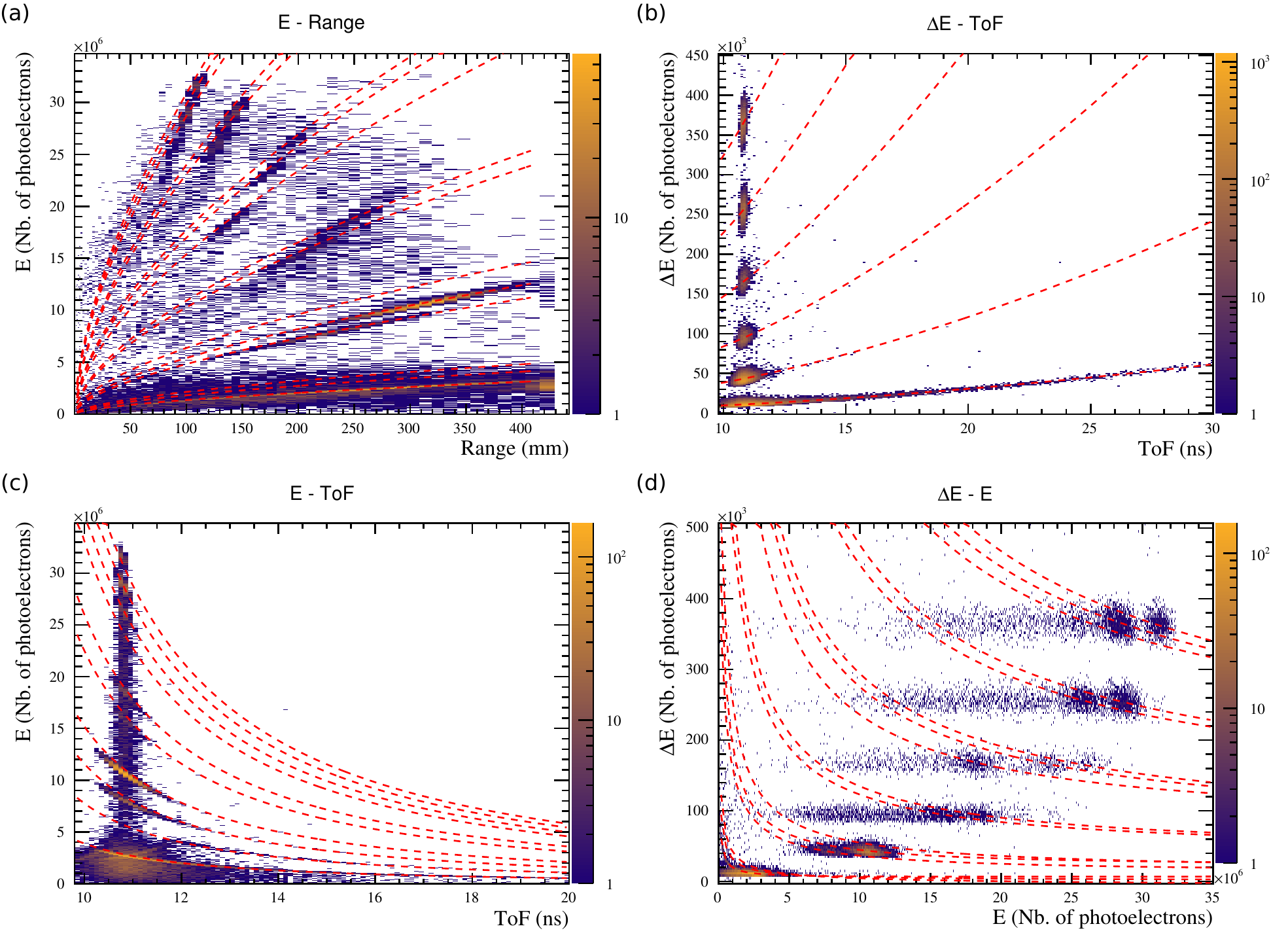}
\centering
\caption{Distributions of (a) E versus range, (b) $\Delta E$ versus ToF, (c) $E$ versus ToF and (d) $\Delta E$ versus the residual energy for a beam energy of 400~MeV/A. Red dashed lines represent the curves used for identification.}
\label{figFits400MeV}
\end{figure*}

Figure~\ref{figFits400MeV} gives an example of the plots of the different identification methods with their associated fitted curves at $\textrm{E}_{beam}=400~\textrm{MeV/A}$. The color scales denote the number of events per bin and the energy is expressed as the number of collected photoelectrons. Dahsed lines represent the curves used for identification.

\begin{figure*}[!ht]
\includegraphics[clip=true, trim=0.2cm 0. 1.cm 0.75cm, width=14.cm]{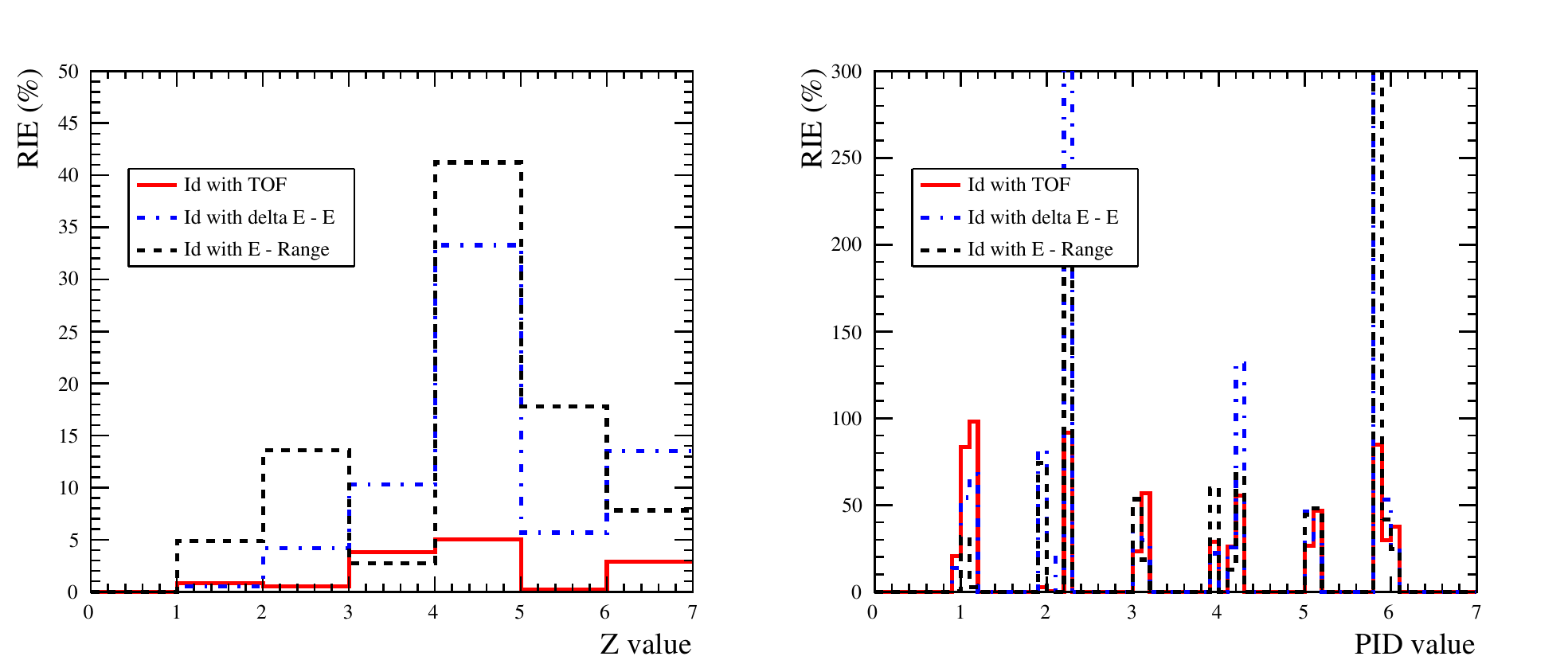}
\centering
\caption{Relative identification errors versus Z and PID value for $\textrm{E}_{beam}=400~\textrm{MeV/A}$. Red solid line, $\Delta E$---$E$---ToF; blue dotted-dashed line, $\Delta E$---$E$ and black dashed line, $\Delta E$---Range method.}
\label{figId400MeV}
\end{figure*}

Fig.~\ref{figId400MeV} presents an example of the relative identification error for the three methods as the charge (Z) and the PID relative identification error. Values higher than 100\% refer to isotopes identified with a higher statistic compared to the generated one. This occurs when the pollution induced by heavier particles that have experienced an inelastic collision in the CsI layer is large compared to the statistic of the generated particles (see section~\ref{discussion} for details).

\begin{table*}[!ht]
\caption{Averaged relative identification errors with respect to the method and beam energy. Maximum value in braces.}
\label{tab2}
\centering
\footnotesize
\renewcommand{\arraystretch}{1.4}
{
\begin{tabularx}{\linewidth}{XXXXXXX}
\hline\hline
Beam energy &  \multicolumn{3}{c}{Z identification error [max.] (\%)}  & \multicolumn{3}{c}{PID identification error [max.] (\%)} \\ \cline{2-4} \cline{5-7}
(MeV/A)     &  $\Delta E$---ToF & $\Delta E$---$E$ & $\Delta E$---Range  &  $\Delta E$---$E$---ToF & $\Delta E$---$E$ & $\Delta E$---Range\\ \hline
95  & 1.1$\pm$0.6 [2.1] & 1.6$\pm$1.5 [4.7] & 1.8$\pm$2.0 [6.0] & 5.5$\pm$4.6 [16.5] & 15.3$\pm$21.2 [86.0]  & 93.9$\pm$221.2 [919.8]\\
200 & 1.3$\pm$0.9 [2.4] & 5.2$\pm$4.8 [15.3] & 5.3$\pm$5.4 [16.4] & 16.0$\pm$13.9 [53.6] & 49.3$\pm$87.8 [339.0] & 31.9$\pm$47.0 [188.3] \\
300 & 1.8$\pm$1.5 [4.2] & 7.1$\pm$6.3 [19.9] & 8.7$\pm$7.6 [24.1] & 41.4$\pm$28.8 [93.3] & 87.1$\pm$206.0 [861.3]  & 37.0$\pm$42.0 [168.0]\\
400 & 2.2$\pm$1.8 [5.0] & 11.3$\pm$10.7 [33.0] & 14.6$\pm$12.9 [41.2] & 44.0$\pm$30.0 [98.1] & 138.5$\pm$275.7 [1132.0] & 71.8$\pm$107.7 [1435.0] \\
\hline\hline
\end{tabularx}
}
\end{table*}

Table~\ref{tab2} summarizes the average relative identification errors for the methods at the different beam energies and gives the highest value in the statistic.
When identifying the charge of the ion, the $\Delta E$---ToF is in average the most efficient method for any beam energy, with an efficiency above 95\% (RIE$\leq$5\%) for any charge. This is mainly due to the good timing resolution and the distance for the time-of-flight measurement while the other methods rely on the energy resolution of the system for this evaluation. When the thickness of the $\Delta E$ stage is well optimized to reduce the amount of inelastic processes and to separate the spots in the $\Delta E$---ToF plot, the energy resolution does not matter so much.
When identifying the PID, i.e. by including the mass of the particles, the $\Delta E$---ToF method cannot achieve at 400~MeV/A an RIE better than 44\% in average and can even attain 98.1\% for tritons. $\Delta E$---$E$ and $E$---Range methods are well above it with 138.5\% and 71.8\% of mean values respectively. For the three methods, the values scale from lower to higher with the increasing beam energy,  except for the $E$---Range method which acts differently at the lowest energy due to the sampling resolution at small ranges. Thus the other two techniques give acceptable results only up to $\textrm{E}_{beam}=95~\textrm{MeV/A}$.

\subsection{Energy evaluation}

\begin{figure}[!ht]
\includegraphics[clip=true, trim=0.2cm 0. 1.5cm 0.75cm, width=\linewidth]{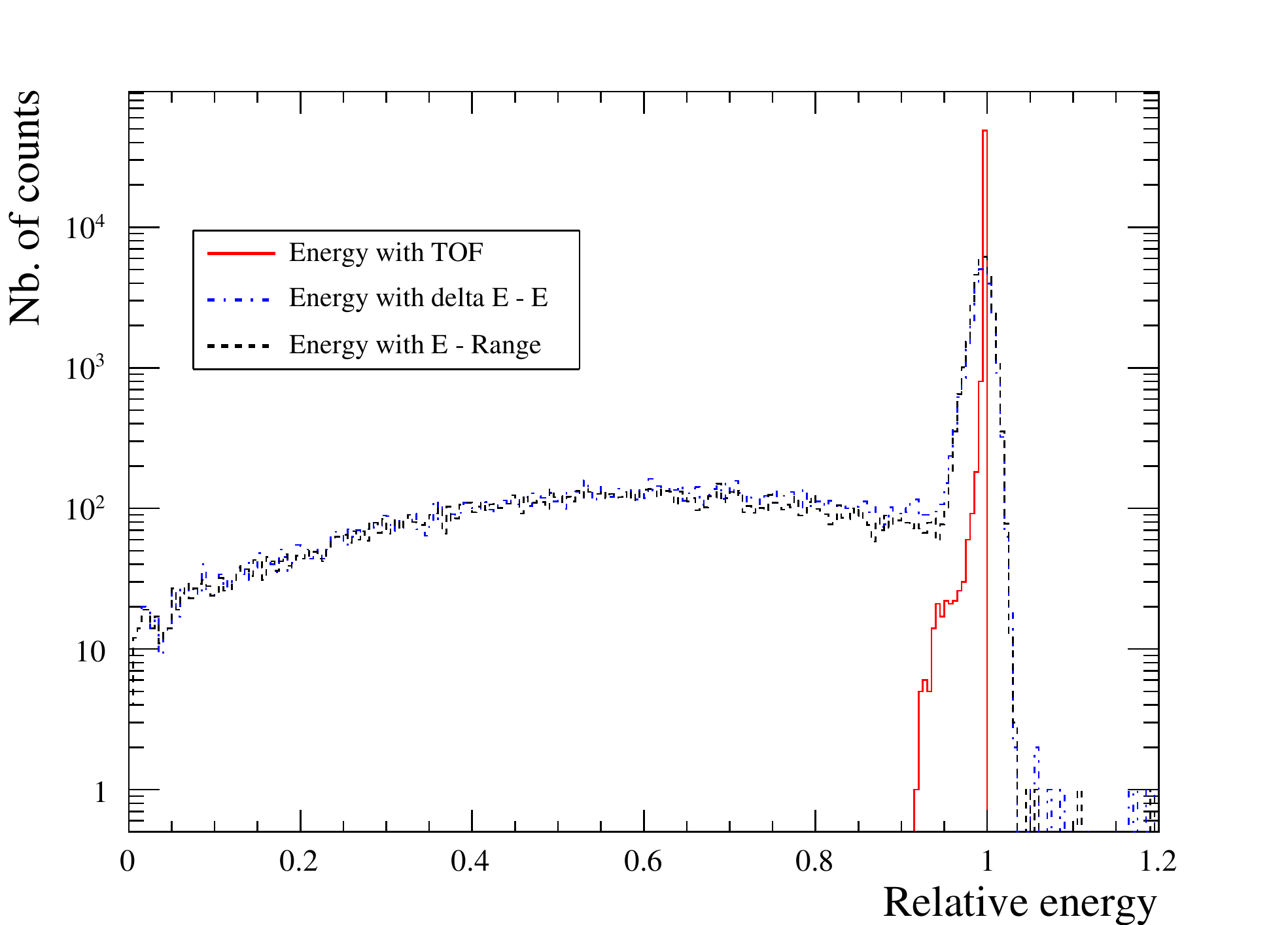}
\centering
\caption{Relative energy distribution compared to the generated one for the three techniques with E$_{beam}$=400~MeV/A. Red solid line, $\Delta E$---ToF; blue dotted-dashed line, $\Delta E$---$E$ and black dashed line, $\Delta E$---Range method.}
\label{figRelEnergy400MeV}
\end{figure}

The relative energy distributions for the three methods are given in Fig.~\ref{figRelEnergy400MeV} for a beam energy of 400~MeV/A. It is clear that even for well identified particles, an amount of energy is lost for a large number of ions, particularly for the $\Delta E$---$E$ and $E$---Range methods.

Table~\ref{tab3} summarizes the different $R_ {trunc}$ values for each technique and beam energy.

\begin{table}[!ht]
\caption{Ratio of ions with truncated measured energy when well identified.}
\label{tab3}
\centering
\footnotesize
\renewcommand{\arraystretch}{1.4}
{
\begin{tabularx}{\linewidth}{XXXX}
\hline\hline
Beam energy &  \multicolumn{3}{c}{$R_ {trunc}$ (\%)} \\ \cline{2-4}
(MeV/A)     &   $\Delta E$---ToF & $\Delta E$---$E$ & $\Delta E$---Range\\ \hline
95  & 18.1 & 11.2 & 10.2 \\
200 & 6.0 & 14.1 & 13.4 \\
300 & 2.6 & 24.9 & 23.4 \\
400 & 1.6 & 38.2 & 32.8 \\
\hline\hline
\end{tabularx}
}
\end{table}

While $R_ {trunc}$ for the $\Delta E$---ToF improves with the beam energy, from 18.1\% to 10.2\%, the other two techniques tends to degrade it drastically. They are both in the same range of values to attain 38.2\% and 32.8\% of particles with the energy truncated, for $\Delta E$---$E$ and $\Delta E$---Range respectively.


\section{Discussion}
\label{discussion}
First, it is interesting to note that the $E$---Range method is in every energy cases the method that gives the highest RIE. This one is well suited for low velocity particles interacting in a gaseous detector, but offers very poor identification performances when used with a dense material and medium to high velocities particles.

To identify the mass of the ion in the given methods, one should measure the residual or total deposited energy. However, the measurement of the deposited energy is often degraded due to inelastic processes (i.e. the nucleus-nucleus collisions) which release a non negligible amount of energy through gamma or neutrons escape and, with a smaller contribution, Q value of the reaction. The lowered measured energy pollutes the identification process of lighter particles by in-between curves data points with strong horizontal lines for the $\Delta E$---$E$ plot or vertical ones for the $E$---ToF plot. Unfortunately, even if the cross-section of nuclear interactions is low compared to electromagnetic processes, the long traveling path of the particles in a rather large detector increases tremendously the probability of such reactions. This effect scales then with the particles kinetic energy worsening the RIE when increasing the beam energy.

\begin{figure}[!ht]
\includegraphics[clip=true, trim=0.cm 0. 10.cm 7.8cm, width=\linewidth]{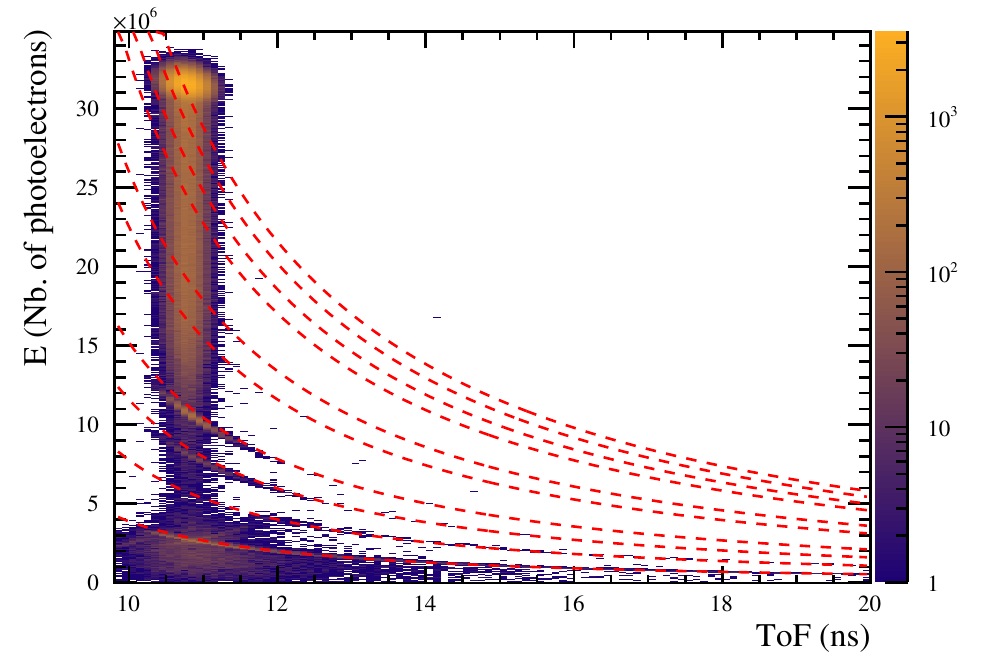}
\includegraphics[clip=true, trim=0.cm 0. 10.cm 7.8cm, width=\linewidth]{400MeV_fits.pdf}
\centering
\caption{Comparison of the $E$---ToF plots with (top) and without (bottom) the beam particles at $\textrm{E}_{beam}=400~\textrm{MeV/A}$.}
\label{figCompCarbons}
\end{figure}

As stated in section~\ref{MatandMeth}, the different method plots did not take into account the beam particles interacting directly with the detection system. The amount of beam particles encountering no fragmentation processes in the target represents approx. 88\% of the cases at 400~MeV/A with a 5~mm diameter target. They can then most likely have inelastic processes in the detector itself due to its dimensions, degrading the particle identification as mentioned previously. Figure~\ref{figCompCarbons} shows a comparison of the $E$---ToF plot with and without the beam particles at 400~MeV/A. It is clear that including beam particles in the identification process would result in artificially degrading them and lead us to a different interpretation of the results, while the goal is to compare identification methods and not detection systems.

Even when well identified, a particle might have its measured energy degraded. It can be attributed to two major effects: inelastic collisions mainly in the $E$ stage and detector geometrical effects. Inelastic collisions may not be sufficient to misidentify the particle but the loss of energy by neutrons or gamma rays can be enough to truncate the measured energy. In the case of geometrical effects, the particles can escape by the sides due to lateral scattering, or by the back due to a high velocity (for light particles only). This last can be avoided by using a larger detector but would then increase the cost and the number of channels of the system.

The two techniques, $\Delta E$---$E$ and the $\Delta E$---Range, have a very noxious incidence on the energy measurement. At $\textrm{E}_{beam}=400~\textrm{MeV/A}$, the energy of more than 30\% of the ions cannot be evaluated precisely, regardless of the energy resolution. In the end, this will result in larger error bars of the production cross sections relative to the energy.

In the case of the $\Delta E$---ToF technique, one would think that $R_ {trunc}$ would increase with the beam energy. However, we can observe the opposite. This effect comes from the method to evaluate the number of ions with a truncated energy. The energy resolution evaluated by ToF scales with the beam energy (see eq.~\ref{eqEnResol} in the case of non relativistic particles), a lower beam energy gives a better energy resolution. For the lowest beam energy, the evaluation of the FWHM of the peak centered at 1 gives a very small value. Then, more ions are included to be with a truncated energy outside of this peak. When increasing the beam energy, the FWHM peak value becomes larger and less ions are included in the R$_{trunc}$ value.

\begin{eqnarray}
\label{eqEnResol}
 \frac{\sigma_E}{E}=\frac{2\sigma_t}{ToF}, 
\quad\textrm{with}\quad\frac{1}{ToF}\propto E,
\end{eqnarray}
where $\sigma_t$ is the coincidence time variance and taken as a constant.

It is then hard to tell to which lower limit to include ions in the R$_{trunc}$ value, that is why an arbitrary value of -5$\sigma$ values was used regardless of the method.

The same effect is however hidden in the other two techniques, resulting in the oppposite effect due to a worsening intrinsic energy resolution, following an E$^{-1/2}$ trend, artificially improving results at low energy.
~\\

Finally, none of the presented methods is able to identify the particle masses with a sufficiently good efficiency for a beam energy above 95~MeV/A using only scintillating crystals. As for different types of detector, the goal would always be to maintain an inelastic collisions rate as low as possible in order to have the smallest error on the measurement of the deposited energy.


\section{Conclusion}

In this work, we have performed Monte Carlo simulations to test several particle identification techniques. To be used in multi-fragmentation experiments associated to carbon beams, we tested techniques based on the measurement of the partial, the total or residual deposited energy, the range as well as the time-of-flight of the particles. A detector composed of multiple layers of scintillating inorganic crystal offers a good flexibility for testing the identification techniques. The best of them, the $\Delta E$---$E$---ToF method, can only reach a PID RIE of 5.5\% and 44\% for a beam energy of 95~MeV/A and 400~MeV/A respectively. The other two methods give worse results. The use of the $\Delta E$---ToF method to obtain the charge of the particles associated to a deflecting magnet is a more efficient method for mass measurements~\cite{Pleskac2012130,SAMURAI}, not without an increase in the development costs.

\section*{References}

\bibliographystyle{elsarticle-num}
\bibliography{science}

\end{document}